\documentclass{PoS}

\title{The broadband emission properties of AGN jets}

\ShortTitle{The broadband emission properties of AGN jets}

\author{\speaker{C.~S.~Chang}$^a$, E.~Ros$^{b,a}$, M.~Kadler$^{c,d}$, M.~B\"ock$^c$, J.~Wilms$^c$, M.~F.~Aller$^e$, H.~D.~Aller$^e$, L.~Fuhrmann$^a$, E.~Angelakis$^a$, and I.~Nestoras$^a$ \\
        \llap{$^a$}Max-Planck-Institut f\"ur Radioastronomie, Bonn, Germany\\       	
        \llap{$^b$}Universitat de Val\`encia, Spain \\
        \llap{$^c$}Dr. Remeis-Sterwarte \& ECAP, Germany \\ 
        \llap{$^d$}CRESST/NASA GSFC \& USRA, USA \\
        \llap{$^e$}University of Michigan, USA \\
        E-mail: \email{cschang@mpifr.de}}

\abstract{The origin of the high-energy emission of blazars is still a matter of debate. To investigate the emission mechanism of extragalactic outflows and to pin down the location of the emission, we have constructed a broadband spectral energy distribution (SED) database covering from the radio to the gamma-ray band for the complete MOJAVE sample, which consists of 135 relativistically beamed AGN with well-studied parsec-scale jets. Typically, the broadband SEDs of blazars shows a double-humped profile. It is believed that the lower-energy hump is due to synchrotron emission from the radio jet, and the higher-energy hump is generated by i) inverse-Compton upscattered seed photons (leptonic), ii) proton-induced shower (hadronic). Combining the results of high-resolution VLBI observations and the $\gamma$-ray properties of the MOJAVE sources, we attempt to reveal the origin of the high-energy emission in relativistic jets, and search for correlations between VLBI and high-energy properties. }

\FullConference{10th European VLBI Network Symposium and EVN Users Meeting: VLBI and the new generation of radio arrays\\
		September 20-24, 2010\\
		Manchester UK}

\begin{document}

\section{Introduction}

Active Galactic Nuclei (AGN) are among the most energetic objects in the Universe. AGN dominate the extragalactic high-energy sky, and are very active at all wavelengths from radio to $\gamma$-rays. According to the unified model of AGN, it is believed that a super massive black hole is located in the center of host galaxy, and it fuels the whole system with matter accreted around the central engine. In the radio-loud scheme, an energetic jet is launched in the vicinity of the central engine following the warped magnetic field from the pole direction of the accretion disk. Blazars\footnote{Usually, we use the term blazar to refer to BL Lac objects and flat spectrum radio quasars (FSRQ).} are AGN whose jets are pointing toward us, and dominate the radio and the high-energy sky. After the beginning of operations of the Large Area Telescope (LAT) on-board the \textit{Fermi} $\gamma$-ray Space telescope in mid 2008, \textit{Fermi}/LAT detected 709 AGN in the first 11 months, and 85\% of them being blazars \cite{abdo10a}.  The high-energy emission location in blazar systems is not yet well-understood. By using the very long baseline interferometry (VLBI) technique, we are able to resolve jet structure and trace component ejection at milli-arcsecond scales \cite{lister09}. Combining the VLBI with the high-energy observatories (e.g., \textit{Fermi}/LAT), we might be able to probe the location of the high-energy emission. 

The \textbf{M}onitoring \textbf{O}f \textbf{J}ets in \textbf{A}ctive Galactic Nuclei with \textbf{V}LBA \textbf{E}xperiments (MOJAVE) program has been monitoring a radio-selected sample since the mid 1990s. The sample contains mostly blazars due to the selection criteria\footnote{(1) J2000.0 declination$\leq-$20$^{\circ}$; (2) galactic latitude |b|$\leq$2.5$^{\circ}$; (3) 15\,GHz VLBI flux density$\geq$1.5\,Jy.} used \cite{lister09a}. In the \textit{Fermi} one-year AGN catalog \cite{abdo10b}, 63\% of the MOJAVE sources were detected. It was found that the LAT-detected MOJAVE sources have higher brightness temperature and higher Doppler-boosting factors than the non-detected ones \cite{kovalev09}, and the authors suggested that the parsec-scale radio core is likely to be the location of the radio and the $\gamma$-ray flares. It was also reported that the $\gamma$-ray bright quasars have faster jets \cite{lister09b}. 

In order to investigate the relation between the parsec-scale jets and high-energy emission, we are studying the broadband spectral energy distribution (SED) from the radio to the $\gamma$-ray of the MOJAVE sample. By comparing the SED properties of the statistical-complete MOJAVE sources with the VLBI parameters, we want to further understand the physical mechanisms ongoing in blazar jets.

\section{The Project}
We constructed the broadband SED catalog of the complete sample of 135 MOJAVE sources using simultaneous observations from the radio to the $\gamma$-ray band \cite{chang10a, chang10b}. In the radio band, we use: i) data with the 26-meter radio telescope at the University of Michigan Radio Astronomy Observatory (UMRAO) \cite{aller85, aller03}; ii) data obtained with the \textit{\textbf{F}ermi}-\textbf{G}ST \textbf{A}GN \textbf{M}ulti-frequency \textbf{M}onitoring \textbf{A}lliance (FGAMMA) program with the Effelsberg 100-m telescope, the IRAM/Pico Veleta 30-m telescope, and the APEX 12-m telescope. In the optical band, we use the \textit{Swift} UV-Optical Telescope (UVOT) observations. In the X-ray band, we use the \textit{Swift} X-ray Telescope (XRT) and the Burst Alert Telescope (BAT) results. In the $\gamma$-ray band, we use the \textit{Fermi}/LAT one-year catalog results for the 85 sources which were detected in the catalog \cite{abdo10b}. For the remaining 50 sources, we used the upper-limits of \textit{Fermi}/LAT to be published in B\"ock et al. (in prep.).  We also included the historical data from the NASA/IPAC Extragalactic Database\footnote{\texttt{http://nedwww.ipac.caltech.edu/}}, which, among others, include data with the EGRET\footnote{\textbf{E}nergetic \textbf{G}amma \textbf{R}ay \textbf{E}xperiment \textbf{T}elescope} onboard \textit{CGRO}\footnote{\textbf{C}ompton \textbf{G}amma \textbf{R}ay \textbf{O}bservatory}. For details of data acquisition, see Chang et al. \cite{chang10a, chang10b}.

In order to model the SEDs, we applied 2--4 degree polynomial fits to the two humps of the SED for each MOJAVE source as a first approach, and derived the peak values in the frequency and in the energy domain.


\section{Results}

\begin{figure*}[Ht]
\centering
 \includegraphics[width=0.8\textwidth]{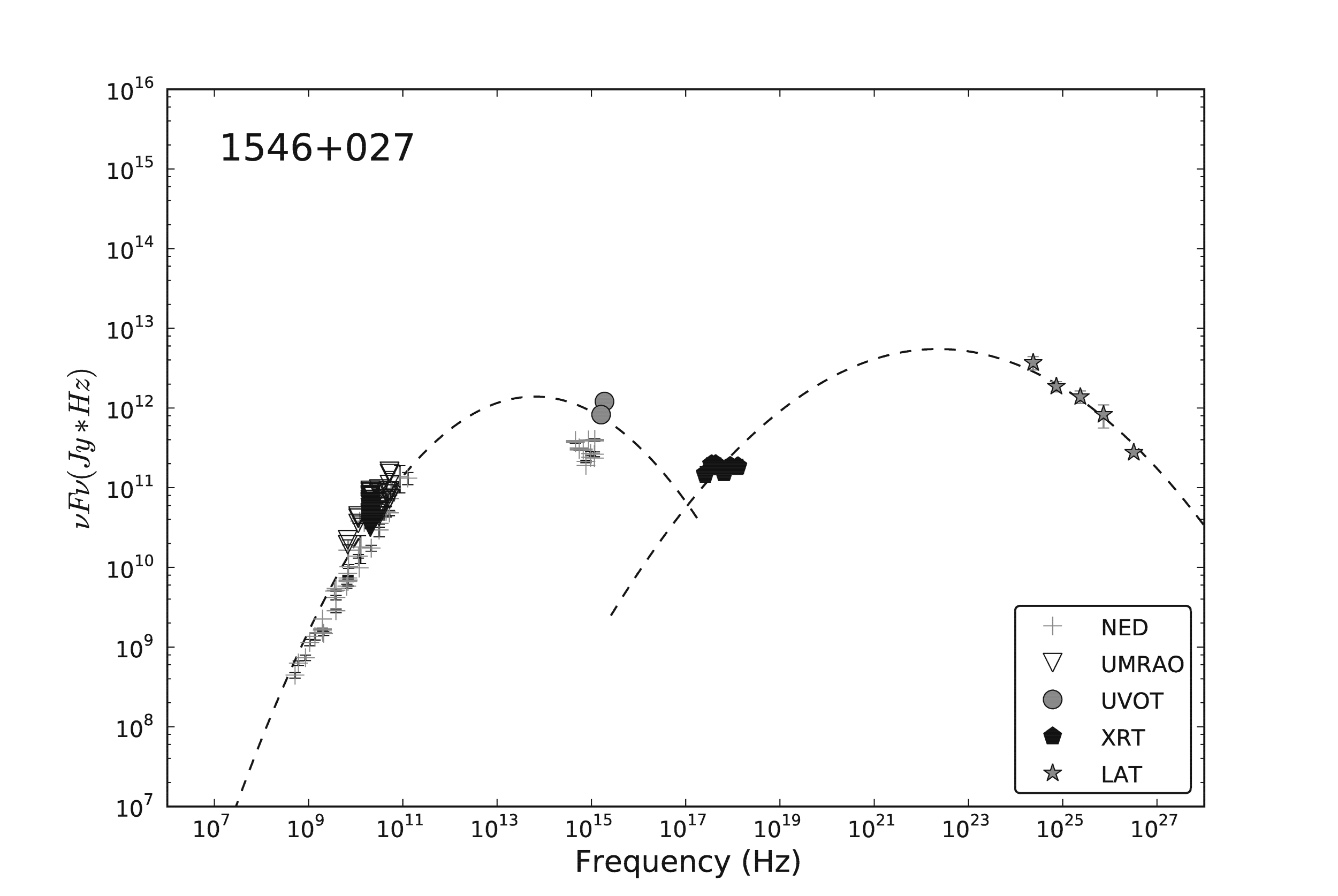}
\caption{The broadband SED of the MOJAVE source 1546$+$027. }
\label{fig:1SED}
\end{figure*}

Figure \ref{fig:1SED} shows the broadband SED of the MOJAVE source 1546$+$027. The frequency coverage of the SED is reasonably good, see also Chang (2010) for details \cite{chang10a, chang10b}. As shown in Figure \ref{fig:1SED}, two second-order polynomial fits describe the double-humped SED reasonably good. From the fits, we estimated the peak values of the lower-energy and the higher-energy hump for most of the sources. $\nu_{\mathrm{peak}}^{\mathrm{low}}$, $\nu^{\mathrm{high}}_{\mathrm{peak}}$, $\nu$F$\nu^{\mathrm{low}}_{\mathrm{peak}}$, $\nu$F$\nu^{\mathrm{high}}_{\mathrm{peak}}$, except for few sources with poorer frequency coverage at the higher energy range. We study the distributions and correlations of the four SED parameters for the different classes of AGN in our sample: quasars, BL Lac Objects, and radio galaxies. We found that the distribution of $\nu_{\mathrm{peak}}^{\mathrm{low}}$ locates at a similar range for three classes of AGN. However, the ranges of the distribution of $\nu^{\mathrm{high}}_{\mathrm{peak}}$ differ: (1) the quasars occupy a much wider range compared to the other two categories, as well as compared to the distribution of $\nu_{\mathrm{peak}}^{\mathrm{low}}$ for quasars; (2) the BL Lac objects occupy higher range of $\nu_{\mathrm{peak}}^{\mathrm{low}}$ than the radio galaxies. While investigating the relations between apparent speed $\beta_{\mathrm{app}}$ and $\nu^{\mathrm{high}}_{\mathrm{peak}}$, we found that there is no source with high apparent speed  having low values of $\nu^{\mathrm{high}}_{\mathrm{peak}}$, which might suggest a relationship between the two parameters. 


\section{Future Work}
The broadband SED catalog of the MOJAVE sample has been constructed \cite{chang10b}. Currently, we are studying the distributions of the SED parameters, as well as the correlations with respect to the radio (VLBI), optical, X-ray, and $\gamma$-ray properties. Present physical SED models can describe the higher energy range well, however, the fitting results are not satisfactory in the radio band. The aims of our correlation studies are to understand the nature of blazar emissions, and also to find out possible linked parameters, which will help to improve blazar SED models. The physical models are essential to understand the local conditions of blazar jets, such as magnetic field intensity, jet composition, bulk flow, etc. The results of the correlation study will be presented in a forthcoming publication (Chang et al. in prep.).



\begin{thebibliography}{99}


\bibitem{abdo10a} A.A.~Abdo, M.~Ackermann, M.~Ajello et al., \textit{Fermi Large Area Telescope First Source Catalog}, \textit{ApJS} \textbf{188} (2010) 405

\bibitem{lister09} M.L.~Lister, M.H.~Cohen, D.C.~Homan et al., \textit{MOJAVE: Monitoring of Jets in Active Galactic Nuclei with VLBA Experiments. VI. Kinematics Analysis of a Complete Sample of Blazar Jets}, \textit{AJ} \textbf{138} (2009) 1874--1892

\bibitem{lister09a} M.L.~Lister, H.D.~Aller, M.F.~Aller et al., \textit{MOJAVE: Monitoring of Jets in Active Galactic Nuclei with VLBA Experiments. V. Multi-Epoch VLBA Images}, \textit{AJ} \textbf{137} (2009) 3718--3729

\bibitem{abdo10b} A.A.~Abdo, M.~Ackermann, M.~Ajello et al., \textit{The First Catalog of Active Galactic Nuclei Detected by the Fermi Large Area Telescope}, \textit{ApJ} \textbf{715} (2010) 429--457

\bibitem{kovalev09} Y.Y.~Kovalev, H.D.~Aller, M.F.~Aller et al., \textit{The Relation Between AGN Gamma-Ray Emission and Parsec-Scale Radio Jets}, \textit{ApJ} \textbf{696} (2009) L17--L21

\bibitem{lister09b} M.L.~Lister, D.C.~Homan, M.~Kadler et al., \textit{A Connection Between Apparent VLBA Jet Speeds and Initial Active Galactic Nucleus Detections Made by the Fermi Gamma-Ray Observatory}, \textit{ApJ} \textbf{696} (2009) L22--L26

\bibitem{chang10a} C.S.~Chang, E.~Ros, M.~Kadler et al, \textit{The Broadband Spectral Energy Distribution of the MOJAVE Sample}, \textit{Proceedings of the Workshop "Fermi meets Jansky - AGN in Radio and Gamma-Rays", Savolainen, T., Ros, E., Porcas, R.W. \& Zensus, J.A. (eds.), MPIfR, Bonn, June 21-23 2010}

\bibitem{aller85} H.D.~Aller, M.F.~Aller, G.E.~Latimer et al., \textit{Spectra and linear polarizations of extragalactic variable sources at centimeter wavelengths}, \textit{ApJS} \textbf{59} (1985) 513--768 

\bibitem{aller03} M.F.~Aller, H.D.~Aller, P.A.~Hughes et al., \textit{Pearson-Readhead Survey Sources. II. The Long-Term Centimeter-Band Total Flux and Linear Polarization Properties of a Complete Radio Sample}, \textit{ApJ} \textbf{586} (2003) 33--51

%

\bibitem{chang10b} C.S.~Chang, \textit{Active Galactic Nuclei throughout the Spectrum: M\,87, PKS\,2052$-$47, and the MOJAVE Sample}, \textit{PhD thesis}, University of Cologne (2010), Chapter 4

\end{thebibliography}
\end{document}